# On a Flywheel-Based Regenerative Braking System for Regenerative Energy Recovery

Tai-Ran Hsu, ASME Fellow
Professor and Chair
Department of Mechanical Engineering
San Jose State University
San Jose, CA 95192

ABSTRACT

This paper presents a unique flywheel-based regenerative energy recovery, storage and release system developed at the author's laboratory. It can recover and store regenerative energy produced by braking a motion generator with intermittent rotary velocity such as the rotor of a wind turbogenerator subject to intermittent intake wind and the axels of electric and hybrid gas-electric vehicles during frequent coasting and braking. Releasing of the stored regenerative energy in the flywheel is converted to electricity by the attached alternator. A proof-of-concept prototype called the SJSU-RBS was designed, built and tested by author's students with able assistance of a technical staff in his school.

## I. INTRODUCTION

The present research involves the design, construction and testing of a flywheel-based regenerative braking system (RBS), the SJSU-RBS. This particular RBS can store the kinetic energy produced by intermittent energy sources otherwise would be lost because the recovered regenerative energy by these sources is often too small to be saved. This unique regenerative braking system (RBS) allows the recovered regenerative energy to be converted into electric energy by an integrated flywheel/alternator unit.

Major components in the SJSU-RBS is presented in Figure 1, in which the "rotary motion generator" may be the spinning axel of an electric vehicle during coasting and braking, or an electric motor driven by the power generated by solar photovoltaic cells with fluctuating solar energy intensity, or a wind turbine rotor rotating at variable speeds by intermittent intake wind. The inertia clutch in the system engages the flywheel/alternator unit to the input shaft at a constant or accelerated rotary speed of the motion generator. The same clutch disengages the same unit from the input shaft of the motion generator at a reduced speed or after complete stop turning of this shaft. The disengaged flywheel/alternator unit, though spins at reduced speed, can continue to produce electricity. A "speed boosting" device such as an epicyclic gear train with a combination of "sun," "satellite" and "rim" gears is introduced in the RBS to boost the spinning speed of the flywheel for maximum storage of the kinetic energy. A unique electric charging system has been developed and it is attached to the SJSU-RBS for "electric energy storage, distribution, and management system" as shown in Figure 1. This system can save any level of electricity generated by the recovered regenerative energy by the RBS using a trickle charge controller and a bank of ultracapacitors. It accumulates the input energy to a level that is high enough to be charged to the principal energy storage of the power plants, or to the batteries of an electric vehicle.

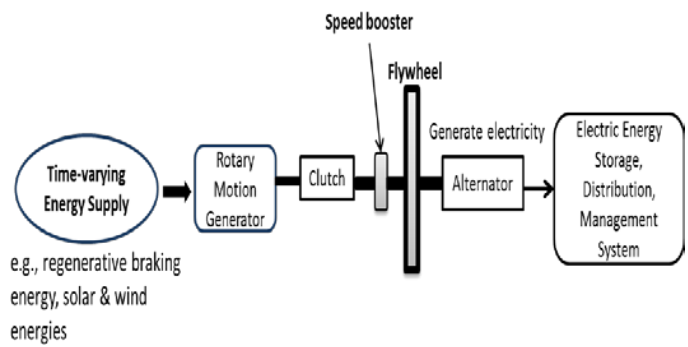

Figure 1 Schematic Diagram of Major Components in the SJSU-RBS

## II. APPLICATIONS OF THE SJSU-RBS IN REGENERATIVE ENERGY RECOVERY AND STORAGE

A. In wind power generation:

A well-known fact in wind power industry is that wind velocity often varies with time. Figure 2(a) shows a recording of significant variation of average wind velocity in a short period of 5 minutes, whereas Figure 2(b) illustrates the frequent fluctuation of wind velocities encountered by wind turbogenerators in the field. Wind turbogenerators are generally designed with a cut-in wind velocity, below which the wind turbine would not be put in operation either because the wind energy below this cut-in velocity is not powerful enough to drive the turbine or the electricity generating unit would not generate sufficient power for storage or transmission to the power grid lines.

The adaption of the SJSU-RBS with major components shown in Figure 1 to a conventional wind turbogenerator unit will enable the flywheel in the RBS to spin to a high speed energized by a small portion of the generated wind power. The flywheel/alternator unit in the RBS releases its stored energy after the wind turbine experiences below the cut-in velocity, or ceases to turn. The free spinning of the flywheel/alternator unit keeps generating electricity until all its stored energy is released. The regenerative braking system in Figure 1 can thus be a significant enhancement to the performance of wind power generation using wind turbogenerators. In theory, the amount of total wind power that can be generated by a horizontal axis wind turbogenerator is:

$$P_{total} = \frac{1}{2} \rho A V^3 (c_b) \qquad (1)$$

where ρ is mass density of the wind in kg/m³, A is the rotor swept area in m², V is the wind speed in m/s and $c_b$ is the Betz coefficient.

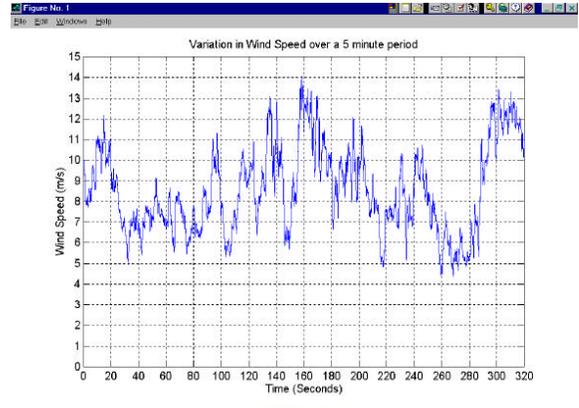

(a) Recorded wind velocity changes in 5 minutes (http://crestdl.lboro.ac.Uk/outside/studyWith Crest/studyNotes/wp01snv1.pdf)

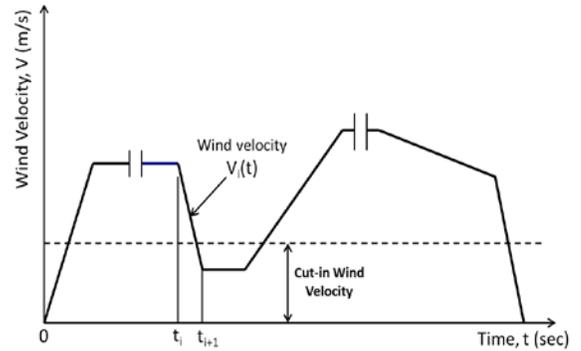

(b) Wind velocity change with time

Figure 2 Fluctuation of Wind Velocity for Wind Power Generation

The Betz coefficient in Equation (1) is obtained by $c_b = \dfrac{(1+V_r)(1-V_r^2)}{2}$, in which $V_r = V_{out}/V_{in}$, with $V_{in}$ and $V_{out}$ to be the respective wind velocities in front and behind the sweeping wind turbine blades. Numerical value of the Betz factor is less than 0.59.

The recoverable regenerative energy for a wind turbogenerator using the SJSU-RBS is a fraction of the available wind power in Equation (1) with decreasing wind speeds $V_i(t)$. For the period between time $t_i$ and $t_{i+1}$ in Figure 2(b), the recoverable regenerative energy can be computed by the following expression:

$$P_{regen\ i} = \frac{1}{2} \eta\, c_b\, \rho A \int_{t_i}^{t_{i+1}} V_i(t)^3\, dt \qquad (2)$$

where $V_i(t)$ is the function describing the variation of the measured wind velocity in the interval between $t_{i+1}$ and $t_i$ as illustrated in Figure 2(b). The factor η in Equation (2) accounts for the loss of the recoverable energy due to the friction of mechanical components in the SJSU-RBS determined by experiments.

B. In power output of electric and gas-electric hybrid vehicles:

Many view electric vehicles (EVs) to be the ultimate means of sustainable ground transportation. Indeed, relentless efforts have been made by the automotive industry to design and produce energy efficient EVs long before the turn of the century. A disappointing fact, however, is that the market share of EVs remains dismal mainly because of the achievable low cruising speed and short cruising range of these vehicles can be produced at affordable prices with the current technology. The gas-electric hybrid vehicle (HEVs) which is designed to achieve much higher gas mileage and produce low emissions but can otherwise perform in par with most gasoline – powered vehicles are well accepted by consumers in this country. The "rotary motion generator" in Figure 1 of EVs and HEVs is their wheel axels. The current technology uses the spinning wheel axels to drive the motor-converted alternator to generate electricity to charge the batteries of the vehicles during coasting and braking. Generation of electricity ceases after the wheel axels stop turning upon the completion of braking of the vehicle. With the separate flywheel/alternator unit in the SJSU-RBS, generation of electricity continues after the complete braking of the vehicle. Literature is available on the working principles of regenerative braking systems of HEVs, and also on their control of transmission systems and regenerative torques and the drive train [1-6].

The laws of both thermodynamics and vehicle dynamics are used to account for the regenerative energies associated with the coasting and braking of a vehicle at any given time t as expressed in Equation (3) [7]:

$$P_{total}(t) = \frac{d}{dt}[KE(t) + PE(t)] + P_{aero}(t) + P_{tire}(t) + other\ terms \qquad (3)$$

where KE(t) is the kinetic energy of the vehicle during coasting and braking at time t, PE(t) is the potential energy during the same time, $P_{aero}(t)$ is the power required to overcome aerodynamic resistance, and $P_{tire}(t)$ is the power required to overcome the resistance of the tires of the vehicle over the road surface.

We envisage that the potential energy PE(t) in Equation (3) may become significant for vehicles coasting on rolling hills. In such cases, both the kinetic and potential energies may be expressed as:

$$KE(t) = [MV(t)^2]/2 \qquad (4a)$$

$$PE(t) = (Mg)\Delta y \qquad (4b)$$

in which M = the mass of the vehicle with payload, V(t) = the velocity of the vehicle during the coasting or braking at time t, g = gravitation acceleration, and Δy = the slope of the road surface with a "+ve" sign for downward slope and "- ve" sigh for upward slope.

The terms $P_{aero}(t)$ and $P_{tire}(t)$ may be significant factors in recovery of regenerative energy of vehicles, and the "other terms" in Equation (3) account for all other power consumptions in the recovery process, which may include mechanical friction of components in the drive and power transmission systems.

II. FLYWHEELS AS MECHANICAL ENERGY STORAGE DEVICES

A flywheel such as the one illustrated in Figure 1 is a mechanical device that is commonly used to store kinetic energy associated with its rotation at high speed. The stored energy is then released to the intended application such as described in Section II after the supplied energy is either discontinued or reduced in the magnitudes.

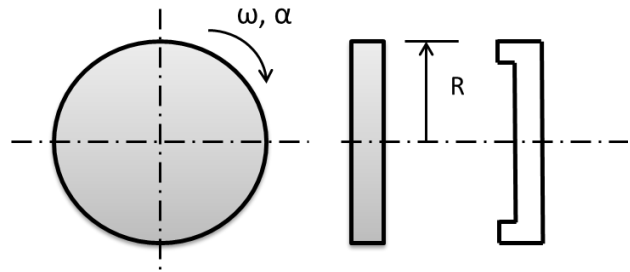

Figure 3 A Spinning Flywheel

Figure 3 shows two popular geometry of flywheels: the uniform cross-section wheels in the middle and the thick-rim flywheels to the right. The kinetic energy that can be stored in a flywheel spinning at an angular velocity ω may be computed by the following expression [8]:

$$KE = \frac{1}{2} I \omega^2 \qquad (5)$$

where I is the mass moment of inertia of the spinning wheel. Equation (6) is used to compute the mass moment of inertia for flywheel with uniform thickness:

$$I = \frac{1}{2} MR^2 \qquad (6)$$

with M and R being the mass and the radius of the wheel respectively.

The angular velocity of the spinning flywheel ω is maintained by applying torques that is equal to $T = I\alpha$ in which α is the angular acceleration of the spinning wheel [8].

III. THE PROOF-OF-CONCEPT PROTOTYPE OF SJSU-RBS

A proof-of-concept prototype that would function like what was described in the forgoing sections was designed and constructed by author's graduate student with able assistance from a technical staff from his school. The "rotary motion generator" that drives the prototype was a dc motor. A flywheel of the thick-rim geometry made of aluminum alloy with outside diameter 11.5 inches and 0.75 inch thick was attached to a turning wheel as illustrated in Figure 4 (a). Figure 4(b) shows the assembled finished product.

As illustrated in Figure 4(a), the flywheel, the alternator and the "sun gear" of the "epicyclic gear train" share a common shaft whereas the satellite and the ring gears in the epicyclic gear train is driven by a thin disk. All these components are encased in a steel "cage" for the protection of the operator and bystanders from probable burst of the flywheel assembly at high spinning speed. The common shaft of the flywheel/alternator assembly is connected to the sprocket that is in turn connected to another sprocket attached to the axel of the "progressive braking system" with a chain link. This braking unit functions like the "inertia clutch" in Figure 1. The progressive braking system consists of a number of pairs of friction metal disks separated by springs. The master hydraulic cylinder provides hydraulic pressure to "push" the friction metal disks from both ends of the cylinder toward the center of the braking unit. The braking of the drive shaft of the dc motor is accomplished by continuous application of hydraulic pressure to the device until the built-in clutch seizes the drive shaft and the sprocket. At which time, the upper sprocket that is attached to the flywheel embark its spinning at a speed that is designed to be 4 times that of the input spinning speed of the input drive shaft from the dc motor that functions as a motion generator.

The SJSU-RBS shown in Figure 4 was placed on a bench-top for proof-of-concept testing. Components involved in this test are shown in Figure 5 with a dc motor driving the input drive shaft together with a data recording and processing unit using a laptop computer. Measured data included the rotary speeds of the flywheel and the driving shaft. The dc motor was driven by 4 12-volt lead-acid batteries (not shown in the figure). Sensors that detect and record the rotary velocities of the flywheel/alternator and the drive shaft were installed. Braking to the drive shaft through the progressive braking device was induced by applying force to the brake arm shown at the left of Figure 5.

Figure 6 illustrates the anticipated variations of input angular velocities of the drive shaft of the dc motor and the induced angular velocities of the flywheel/alternator during bench-top testing. The drive shaft was made to spin to a stable angular velocity $\omega_{sh}$ prior to time $t_0$, at which braking to the progressive braking unit is invoked, and the driving shaft is engaged to the flywheel/alternator assembly by the inertia clutch. Consequently, the rotary velocity of the flywheel/alternator unit increases with further braking of the drive shaft and it reaches the maximum angular velocity $\omega_{max}$ at $t_1$. The two units rotate synchronously from time $t_1$ to $t_2$, at which the drive shaft comes to a complete stop at time $t_2$, at which instant the two units separate, resulting in free spinning of the flywheel. The free-spinning of the flywheel/alternator unit continues until the flywheel releases all its stored energy.

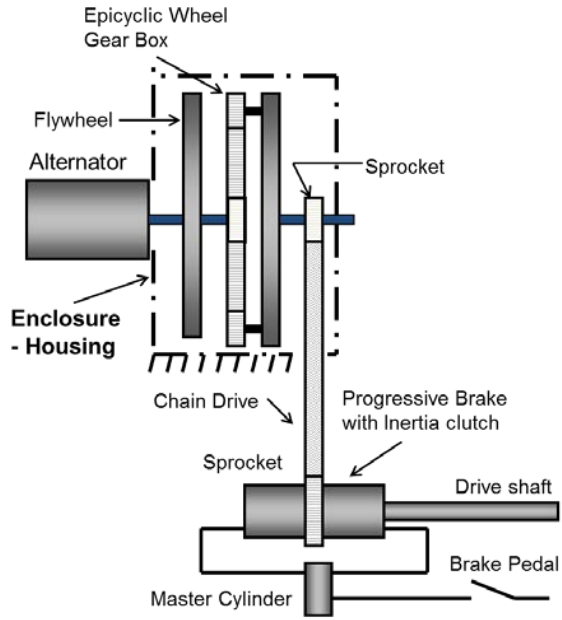

(a) Major components

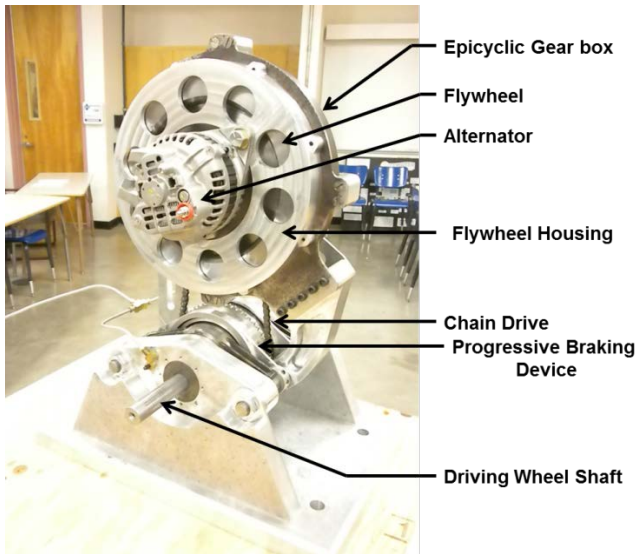

(b) The assembled system

Figure 4 Proof-of-Concept Prototype of the SJSU-RBS

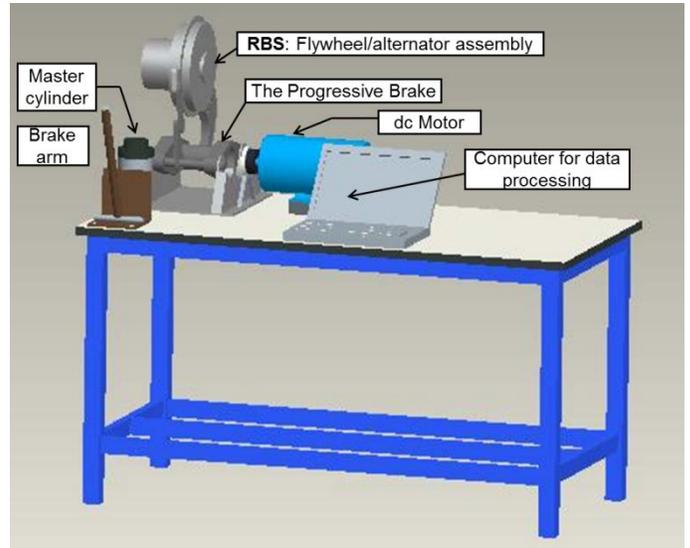

Figure 5 Setup of Equipment for Bench-Top Test of the SJSU-RBS Prototype

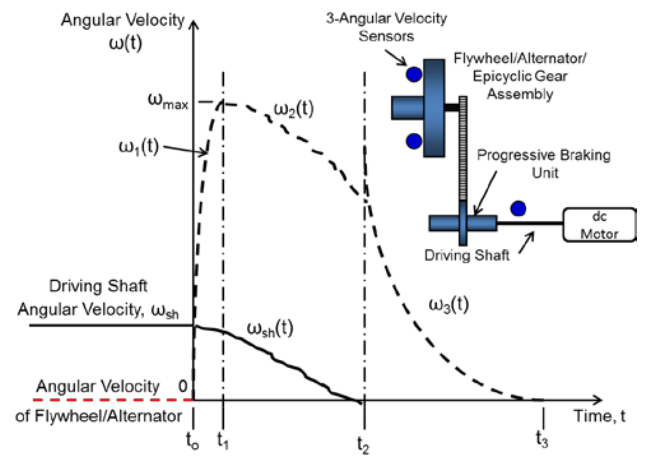

Figure 6 Variation of Angular Velocities of Flywheel and Drive Shaft

The regenerative energy produced by the SJSU-RBS in the bench-top testing may be computed by the following expression:

$$KE_{rec} = \frac{1}{2}I\left\{-\left[\int_{t_0}^{t_1}\omega_1(t)dt\right]^2 + \left[\int_{t_1}^{t_2}\omega_2(t)dt\right]^2 + \left[\int_{t_2}^{t_3}\omega_3(t)dt\right]^2\right\}$$
(7)

where the mass moment of inertia of the flywheel I in Equation (7) is obtained by using Equation (6).

A group of 4 undergraduate students conducted the bench-top testing of the SJSU-RBS in academic year 2011/2012. Three cases were reported on the measured net kinetic energy recovered by the spinning flywheel/alternator unit. Variation of the angular velocities of this unit during the braking period is depicted in Figure 7. One may observe that the diagram shown in Figure 7 is similar to what is shown in Figure 6 with the coincidence of time instants $t_1$ and $t_2$, which indicates that the period of "synchronized motion" of the flywheel and the input drive shaft did not exist in these tests at relatively low rotary velocities. The measured data and the recovered regenerative energies are presented in Table 1.

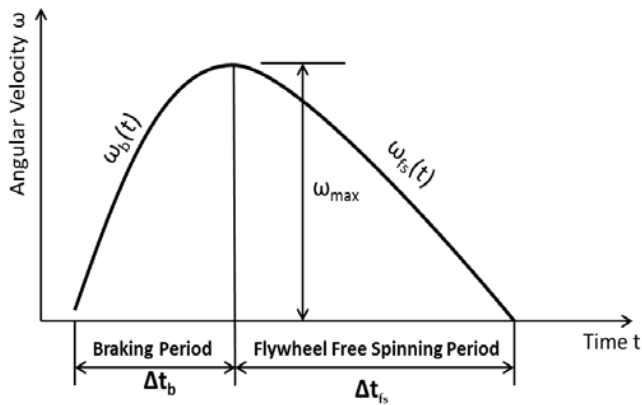

Figure 7 Functions of Spinning Flywheel from Bench-top Testing

Table 1 Measured Angular Velocities of Flywheel in Bench-top Testing of SJSU-RBS
The net kinetic energy produced by the spinning flywheel shown in the right column of Table 1 was obtained by the expression:

$$KE_{rec} = \frac{1}{2} I \left( \int_{\Delta t_{fs}} \omega_{fs}^2(t)\,dt - \int_{\Delta_b} \omega_b^2(t)\,dt \right) \quad (8)$$

in which functions $\omega_{fs}(t)$ and $\omega_b(t)$ are established by curve-fitting of the measured data using a "polynomial function curve fitting" technique.

We may summarize the average net regenerative energies produced by the spinning flywheel as presented in Table 2.

Table 2 Measured Average Regenerative Energy Produced by the SJSU-RBS

| Case No. | Maximum Flywheel Speed (rpm) | Braking Period (sec) | Average Flywheel Free Spin Period (sec) | Average Recovered Energy (J) |
|---|---|---|---|---|
| 1 | 300 | 10 | 27.3 | 593 |
| 2 | 500 | 5 | 29.3 | 1187 |
| 3 | 500 | 10 | 29.3 | 600 |

It appears that more energy is recovered with high flywheel spinning velocities and with shorter braking periods, as expected.

## IV. SUMMARY AND CONCLUDING REMARKS

A new regenerative braking system, the SJSU-RBS was developed with the design, construction and testing of a proof-of-concept prototype. It involves a fast spinning flywheel/alternator unit with a uniquely designed progressive braking system and an epicyclic gear train. This new SJSU-RBS can be readily adapted to power plants driven by renewable energies from intermittent sources such as solar, wind and braking of electric and hybrid gas-electric vehicles during coasting and braking. The SJSU-RBS was proof-tested for its feasibility and practicality for the intended applications.

Despite the success in the preliminary bench-top testing of the prototype of the SJSU-RBS as presented in the paper, a few key technical issues remain unsolved. Issues such as the optimal design of flywheel for maximum net recovery and storage of regenerative energies; quantification of aerodynamic

| Case No | Rotary velocity of flywheel, $\omega_{max}$ (rpm) | Braking period $\Delta t_b$ (sec) | Flywheel free spinning period $\Delta t_{fs}$ (sec) | Net recovered kinetic energy, (J) |
|---|---|---|---|---|
| 1 | 300 | 10 | 28 | 509 |
|   |     | 10 | 24 | 670 |
|   |     | 10 | 30 | 600 |
| 2 | 500 | 5  | 30 | 1181 |
|   |     | 5  | 28 | 1279 |
|   |     | 5  | 30 | 1102 |
| 3 | 500 | 10 | 30 | 695 |
|   |     | 10 | 28 | 500 |
|   |     | 10 | 30 | 605 |

and electromechanical resistance to the free spinning of the flywheel, and the effective and optimal control of the motion of the flywheel and the driving shafts, etc. will have significant effects on the performance of the SJSU-RBS or similar regenerative braking system for maximal recovery of regenerative energies in reality.

Further research on the detailed design and integration of the SJSU-RBS to wind power generating plants and EVs and HEVs for performance enhancements is desirable. The success of such integration will result in great economical returns to the renewable power generation industry. Efficient power generations by renewable energy sources by RBS will make significant contributions to the sustainable development of global economy and well-being of all humankind.

## ACKNOWLEGEMENT

The author is indebted to the valuable assistance that he received from a late colleague and technologist Craig Stauffer at his university. It was with his brilliant idea and superior skill in machining made the production and testing of the SJSU-RBS prototype possible. Financial assistance with a faculty development grant for this project by the College of Engineering at the San Jose State University in academic year 2010/2012 is gratefully acknowledged.

## BIOGRAPHY

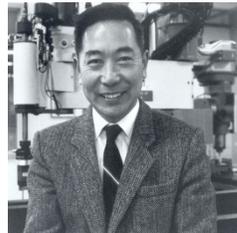

Tai-Ran Hsu, a Fellow of ASME, is currently a Professor and the Chair of the Department of Mechanical Engineering at San Jose State University. He received a BS degree from National Cheng-Kung University in China; MS and Ph.D. degrees from University of New Brunswick and McGill University in Canada respectively. All his degrees were in mechanical engineering. He worked for steam power plant equipment and nuclear industries prior joining the academe. He taught mechanical engineering courses and served as department heads at universities in both Canada and USA. He has published over 120 technical papers in peer reviewed systems and eight books in finite element method in thermomechanics, computer-aided design, and a well-received textbook on microelectromechanical systems (MEMS) design and manufacture. The second edition of the latter book was published in March 2008.